\documentclass[twocolumn,tighten]{aastex631}
\usepackage{graphicx}
\usepackage{gensymb}
\usepackage{amsmath}
\begin{document}
\title{Discovery of a [O III] Emission Shell Around the X-ray Binary CI Cam }

\author[0000-0003-3829-2056]{Robert A.\ Fesen}
\affiliation{6127 Wilder Lab, Department of Physics and Astronomy, Dartmouth College, Hanover, New Hampshire, 03755, USA}

\author[0000-0002-7855-3292]{Marcel Drechsler}
\affiliation{\'Equipe StDr, B{\"a}renstein, Feldstraße 17, 09471 B{\"a}renstein, Germany}

\author{Nicolas Martino}
\affiliation{Various Amateur Observatory Sites, France}

\author{Yann Sainty}
\affiliation{YSTY Astronomy, 54000 Nancy, Lorraine, France}

%\author[0000-0002-3172-965X]{Xavier Strottner}
%\affiliation{\'Equipe StDr, Montfraze, 01370 Saint Etienne Du Bois, France}

%*******************************
\begin{abstract}
We report the serendipitous discovery of a $8' \times 12'$ emission shell brightest in [\ion{O}{3}] centered
on the X-ray binary B[e] star, CI Cameleopardalis (CI Cam). This shell, detected during a survey of optical
emission associated with the Galactic supernova remnant G150.3+4.5, is seen outside but immediately adjacent to the remnant's optical filaments along  its northwestern edge.
Assuming this shell is related to CI Cam's strong winds and transient outbursts, the adoption of CI Cam's Gaia DR3 statistically corrected distance of
4.1$^{+0.3}_{-0.2}$ kpc yields shell dimensions $\simeq$ 9.5 pc $\times$ 14.3 pc. While the distance to the G150.3+4.5 remnant is uncertain, the appearance of CI Cam's emission shell so close to the SNR's optical filaments appears likely to be a chance coincidence.

%with the CI Cam shell a distant background nebulosity to  G150.3+4.5.

%We briefly discuss the nature of this emission shell in regard to CI Cam's evolution.

\end{abstract}
\bigskip

%*******************************
\section{Introduction}
%*******************************

%Evolved supergiant B stars with optical spectra containing forbidden line-emissions called sgB[e] stars exhibit broad high
%excitation lines, narrow low excitation lines, an infrared excess due to
%hot dust, and a two-component stellar
%wind consisting of a normal hot star wind and a slow and dense
%disk-like wind \citep{Lamers1998, Zickgraf1999}.
%In many cases, the stellar photospheres are
%hidden by dense, optically-thick winds. The emission lines
%are from circumstellar wind material, which is often non-spherically symmetric.

The star CI Cameleopardalis 
(CI Cam) was initially cataloged as a Be star 
showing broad H$\alpha$ and \ion{He}{1} emission lines  (MWC~84 = MW~143; \citealt{Merrill1932,Merrill1933a}).
Its optical spectrum was later found to show forbidden lines,
thus classifying it as a B[e] star.
Its spectrum also shows numerous, narrow
[\ion{Fe}{2}] lines, a hot blue continuum, and an infrared excess suggesting
the presence of hot circumstellar dust 
\citep{Downes1984, Allen1973, Miro1995}.
Photometric data indicate CI Cam is a variable with an amplitude of 0.04 mag and a binary with a 19.4 day period based on its He~II lines \citep{Barsukova2023}.

Interest in this star greatly increased following an unusually bright but short transient outburst detected in X-rays on 31 March 1998 (XTE J0421+560) with the Rossi-XTE satellite  recording a peak intensity roughly equal to twice that of the Crab Nebula before rapidly decaying. Its peak optical brightness reached 7.1 mag in R compared to 10.6 mag in its
quiescent state \citep{Barsukova2023}.
Follow-up investigations suggest the primary is a B[e] or sgB[e] star of spectral type B0-B2 with an unseen compact companion either a black hole, a neutron star, or white dwarf.

During the 1998 outburst, radio observations indicated a high-velocity, asymmetric expanding envelope ($\sim14,000$ km s$^{-1}$; \citealt{Miod2004}) followed by a rapid 
deceleration as the shock progressed into dense circumstellar material, the product of the star's strong stellar wind. 
CI Cam is unusual in that it is the only B[e] showing stellar pulsations, and an X-ray behavior unique among X-ray binaries in that its 1998 X-ray outburst was softer compared to X-ray novae. Its outburst kinetic properties of $\sim10^{44-46}$ d$^{2}_{\rm 5~kpc}$ erg \citep{Miod2004, Chugai2010} have also led to the suggestion it may be related to the class of transients known as supernova
imposters \citep{Bartlett2019}.

Here we report the serendipitous discovery of an 
[\ion{O}{3}] emission shell centered on CI~Cam and seen adjacent to optical emission-line filaments associated with the Galactic supernova remnant G150.3+4.5.
%%%%%%%%%%%%%%%%%%%%%%%%%%%%%%%%%%%%%%%%%%%%%%%%%%%%%%%%%%%%%%%%%%%%%
%%% CI Cam Halpha and O3 plots plus color composite
%%%%%%%
\begin{figure*}[ht]
\begin{center}
%\centerlin\includegraphics[angle=0,width=18.5cm]{G82_O3__Radio_Color3.jpg}}
%\includegraphics[angle=0,width=18.0cm]{G82_opt_radio_composite.jpg}
\includegraphics[angle=0,width=18.0cm]{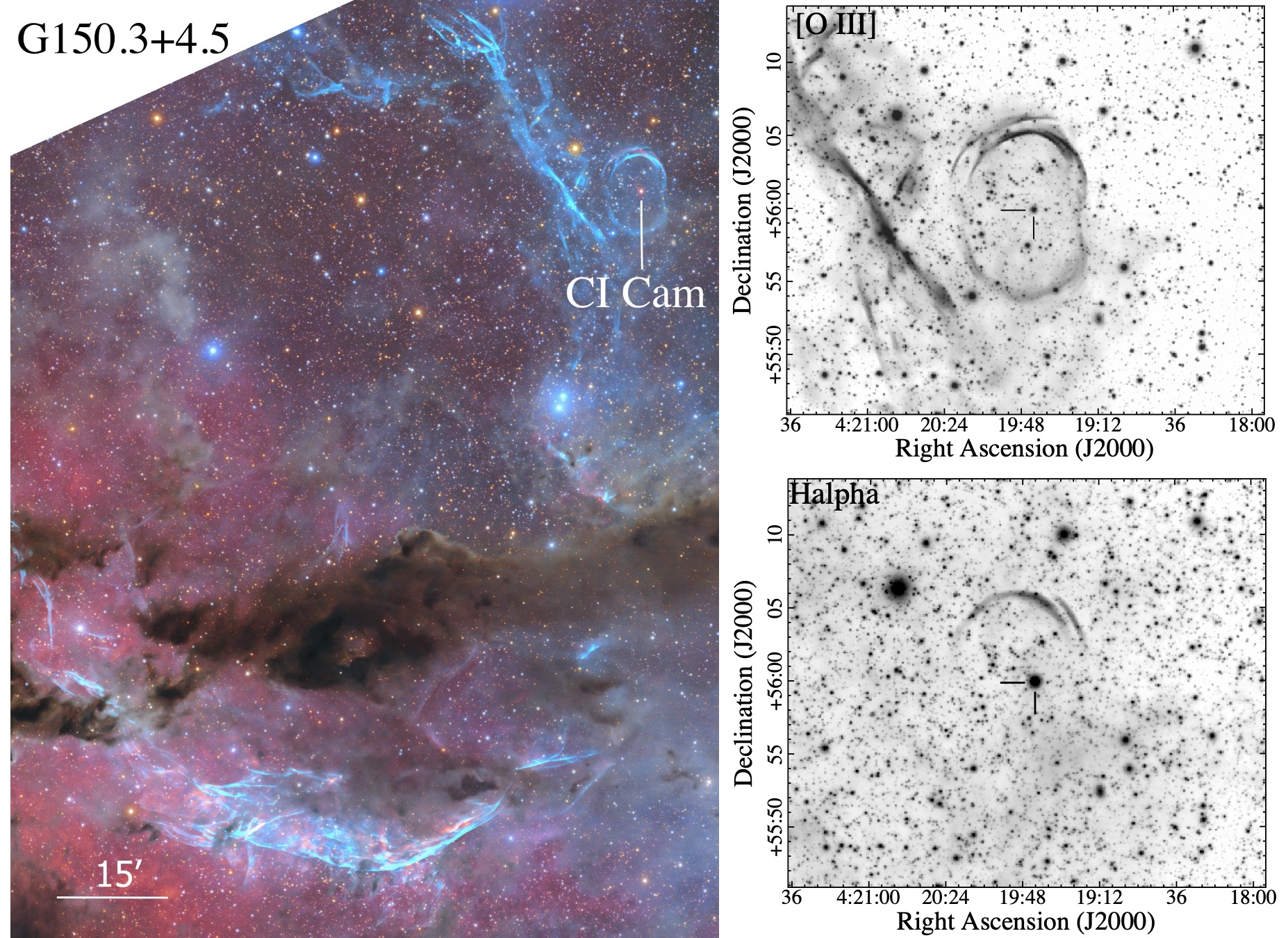}
\includegraphics[angle=0,width=8.85cm]{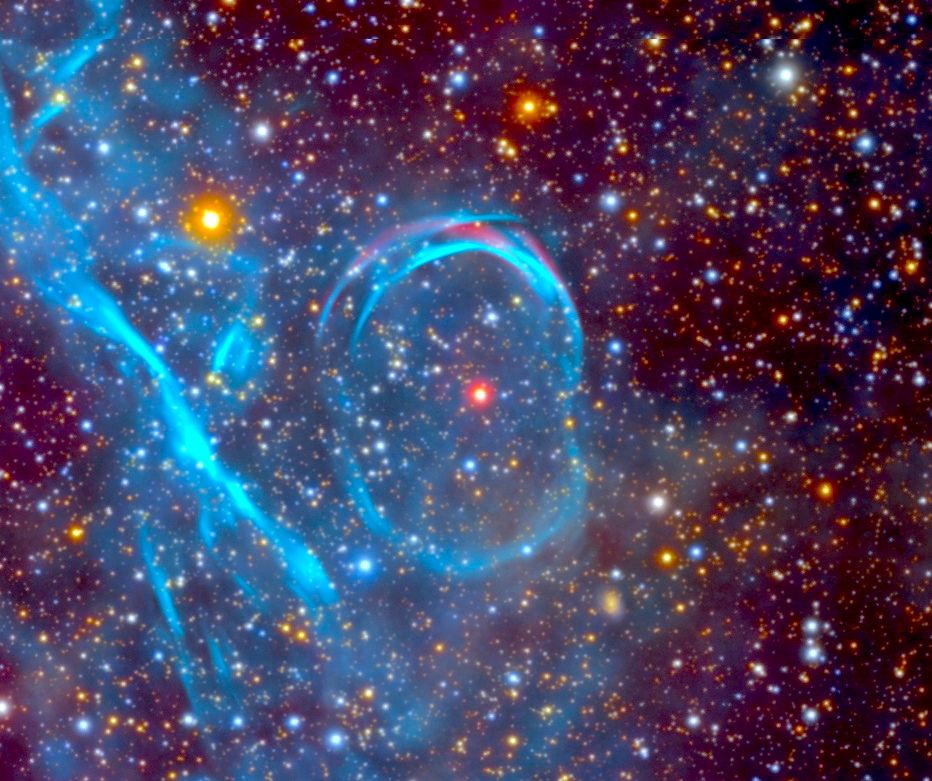}
\includegraphics[angle=0,width=8.68cm]{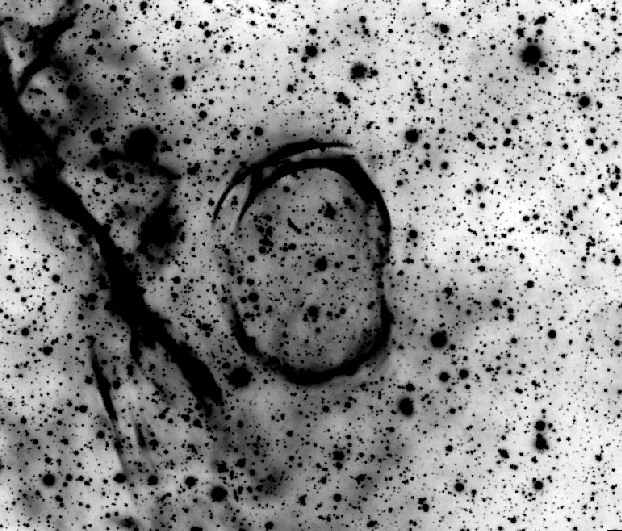} \\
%\centerline{\includegraphics[angle=0,width=15.6cm]{Color_image of CI_v2.jpg}}
\caption{Top Left: [\ion{O}{3}] $\lambda$5007 + H$\alpha$ + RGB color composite of the western portion of the Galactic supernova remnant G150.3+4.5 showing an emission shell centered on CI Cam in upper right corner (from \citealt{Fesen2024})
Top right panels: [\ion{O}{3}]  and 
H$\alpha$ images of the CI Cam emission shell.
Bottom Left: Blowup of color composite image of CI Cam's emission shell. Bottom Right:  Negative  
[\ion{O}{3}] image stretched to show faint emission in the direction of and surrounding CI Cam.   \label{Fig1}  }
\end{center}
\end{figure*}
%%%%%%%%%%%%%%%%%%%%%%%%%%%%%%%%%%%%%%%%%%%%%%%%%%%%%%%%%%%%%%%%%%%%%%

%%%%%%%%%%%%%%%%%%%%%%%%%%%%%%%%%%%%%%
\section{Observations}
%%%%%%%%%%%%%%%%%%%%%%%%%%%%%%%%%%%%%%

As part of a limited optical emission-line survey of Galactic supernova remnants (SNRs) \citep{Fesen2024},
wide-field exposures were obtained of the large  $2.5\degr \times 3\degr$ Galactic supernova remnant G150.3+4.5 
\citep{Gerbrandt2014,Gao2014} using narrow passband emission-line filters along with broad passband continuum filters.
These images were obtained using a Takahaski FSQ-85ED and a Takahaski FSQ-106ED
refractor with CMOS cameras which provided field-of-view of $3.48\degr$ and $2.7\degr$
with image scales of 4.6$''$ and 2.04$''$~pixel$^{-1}$, respectively.
Narrow passband H$\alpha$ and [\ion{O}{3}] filters 
($\sim$ 30 -- 50 \AA \ FWHM) were used.
Total exposure times obtained were: 10.0 hr ($60 \times 600$s) using a 3.0 nm 
[\ion{O}{3}] filter, 13.8 hr ($83 \times 600$)s) using a 
dual H$\alpha$ + [\ion{O}{3}] 5.0 nm filter, and 22.5 hr ($270 \times 300$s) in broadband RGB filters.
These data were obtained over several nights in January and March 2023 at 
Puzieux, Moselle and Moydans, Haute-Alpes in southern France.

%%%%%%%%%%%%%%%%%%%%%%%%%%%%%%%%%%%%%%%%%%%%%%%%%%%%%%%%%%%%%%%%%%
%%%%%%%%%%%%%%%%%%% RESULTS and DISCUSSION %%%%%%%%%%%%%%%%%%%%%%%%%%%%%%%%%%%%% 
%%%%%%%%%%%%%%%%%%%%%%%%%%%%%%%%%%%%%%%%%%%%%%%%%%%%%%%%%%%%%%%%%%

\bigskip

\section{Results and Discussion}

Our imaging of the Galactic SNR G150.3+4.5 revealed a previously unknown emission shell around the high-mass, X-ray binary star CI Cam.
The shell appears immediately adjacent to optical filaments associated with the SNR along its northwestern boundary (see upper left panel in Fig.\ 1). The emission shell is elliptical in shape with outer angular dimensions $\sim 8' \times 12'$.  While the shell's optical emission is dominated by [\ion{O}{3}] $\lambda$5007 line emission, 
a part of an emission arc along the shell's northern section is also visible in  
H$\alpha$ emission (see upper right panels in Fig.\ 1).
The shell is tilted slightly east of north at a position angle (PA)  $\sim 5 - 10$ degrees. Although  much less than the PA = 35$\pm2\degr$ for the star's small dust shell (7.6 mas; \citealt{Thureau2009}), it is close to the PA of 7 degrees for CI Cam's expanding radio emission seen in VLBI data 1 to 306 days following its 1998 outburst \citep{Miod2004}.

The distance to CI Cam has long been uncertain with prior estimates spanning
a range of 1 kpc to $\geq$ 5 kpc \citep{Hynes2002, Robinson2002, Barsukova2006}.
However, Gaia DR3 data, after accounting for statistical corrections, indicates a distance of 4.1$^{+0.3}_{-0.2}$ kpc.
Assuming this emission shell is the result of both CI Cam's strong stellar winds
($\geq 10^{-6} M_{\odot}$ yr$^{-1}$; \citealt{Robinson2002})
and prior transient outbursts like that observed in 1998,  
the emission shell (including the northern partial outer shell) has physical dimensions $\sim$ 9.5 pc $\times$ 14.3 pc. 
These dimensions are comparable to the 5 pc to 40 pc range seen for Galactic and LMC Wolf-Rayet stars
\citep{Chu1999, Hung2021}.

The similarity of 
the CI Cam shell's [\ion{O}{3}] brightness plus its closeness to the SNR's filaments  raises the question of whether or not the two objects might lie at roughly the same distance and be somehow related. 
Unlike G150.3+4.5's optical emission filaments along it southern boundary which show both [\ion{O}{3}] and 
H$\alpha$
emissions, remnant filaments near the CI Cam shell
mainly display only [\ion{O}{3}] emission much like the CI Cam shell itself. 
Strong [\ion{O}{3}] emission is a sign of shock velocities around 90 to 150 km s$^{-1}$. The lack of appreciable 
H$\alpha$ emission in SNRs can signal a shock
progressing through a largely ionized medium, a conclusion consistent with
the diffuse [\ion{O}{3}] emission seen around and seemingly out ahead of the remnant filaments (see bottom panels of Fig.\ 1).

However, there is no emission or structural evidence for any direct physical connection, 
and based on current distance estimates for G150.3+4.5, a connection between CI Cam and the
G150.3+4.5 seems doubtful.
An analysis of the SNR's $\gamma$-ray properties, besides
suggesting it is a dynamically young remnant due to the lack of X-rays, could only constrain its
distance between 0.7 and 4.7 kpc \citep{Devin2020}.
A study of shell-type SNRs with hard $\gamma$-ray spectra suggested
a distance of 1 kpc and a diameter of 48 pc \citep{Zeng2021}, and
a recent investigation of molecular clouds in the direction of
G150.3+4.3 suggests a  remnant distance of around 0.75 kpc leading to remnant dimensions of 33 pc $\times$ 40 pc \citep{Feng2024}. If, however, the remnant’s distance was somehow actually $\sim4$ kpc like that of CI Cam, then its angular diameter of $2.5\degr \times 3.0\degr$ would
imply unlikely dimensions of 180 pc $\times$ 215 pc
ranking it as perhaps the largest and oldest Galactic SNR
known despite its bright [O III] 
filaments indicating relatively high shock velocities.
Consequently, the appearance of the CI Cam shell near
filaments of G150.3+4.5 would seem to be a
chance coincidence.

%with the CI Cam shell a distant background nebulosity to the G150.3+4.5 SNR. 

%Relative emission-line ratios seen in optical spectra of the CI Cam shell could help determine the shock velocity responsible for the shell's emission.

%On the other hand, if G150.3+4.5 
%were the product of multiple SNe, this could help explain both
%a very large size remnant ($\sim 200$ pc) if it were at CI Cam's $\sim$4 kpc distance yet
%display strong high ionization optica lines such as
% [\ion{O}{3}] plus show $\gamma$-ray properties unexpected 
%for such a large and evolved SNR.
%Spectra of both CI Cam's [\ion{O}{3}] emission shell and nearby G150.3+4.5 filaments could offer radial velocity information which could test a SNR-shell association.  

%However, the Galactic region between
%$l$ = 149 -- 153 and $b$ = $1\degr$ -- $5\degr$
%exhibits several arc and partial
%nonthermal radio shells \citep{Kerton2007,Gerbrandt2014}

%\clearpage

\bibliography{ref2}

\end{document}